\documentclass[aps,prl,twocolumn,groupaddress,longbibliography]{revtex4-1}\usepackage{mathrsfs} 
\usepackage{amssymb} 
\usepackage{graphicx,color} 
\usepackage{bbm} 
\usepackage{multirow}
\usepackage{amsmath}
\usepackage{bbm} 
\usepackage{mathrsfs} 
\usepackage{hyperref}
\usepackage[usenames,dvipsnames]{xcolor}
\usepackage{color}
\definecolor{darkgreen}{rgb}{0,0.6,0}
\definecolor{darkblue}{rgb}{0,0,0.6}
\definecolor{darkred}{rgb}{0.6,0,0}
\definecolor{darkpurple}{rgb}{0.5,0,0.5}
\hypersetup{
bookmarksopen=true,
pdftitle="",
pdfauthor="", 
pdftoolbar=false, % toolbar hidden
pdfstartview={FitH},		% fits the width of the page to the window
pdfmenubar=true,			%menubar shown
pdfhighlight=/O,			%effect of clicking on a link
colorlinks=true,			%couleurs sur les liens hypertextes
urlcolor=darkblue,
citecolor=darkblue,		%couleur des liens pour les citations
linkcolor=darkblue}
\newcommand{\beq}{\begin{equation}} \newcommand{\eeq}{\end{equation}}
\newcommand{\bal}{\begin{align}} \newcommand{\eal}{\end{align}}

\newcommand{\argc}[1]{\left[#1\right]}

\newcommand{\argp}[1]{\left(#1\right)}

\renewcommand{\phi}{\varphi}

\newcommand{\eps}{\epsilon}

\newcommand{\tw}{t_{\rm w}}

\let\s=\sigma 

\begin{document}

\title{Rejuvenation and Memory Effects in a Structural Glass}

\author{Camille Scalliet}

\author{Ludovic Berthier}

\affiliation{Laboratoire Charles Coulomb (L2C), Universit\'e de Montpellier, CNRS, Montpellier, France}

\date{\today}

\begin{abstract}
We show numerically that a three-dimensional model for structural glass displays aging, rejuvenation and memory effects when submitted to a temperature cycle. 
These effects indicate that the free energy landscape of structural glasses may possess the complex hierarchical structure that characterize materials such as spin and polymer glasses.
We use the theoretical concept of marginal stability to interpret our results, and explain in which physical conditions a complex aging dynamics can emerge in dense supercooled liquids, paving the way for future experimental studies of complex aging dynamics in colloidal and granular glasses.   
\end{abstract}

\maketitle

The behavior of many disordered materials is dominated by their failure to reach equilibrium, leading to extremely slow relaxations, non-linear responses, and time-dependent behavior. This aging behavior is observed in a broad variety of condensed-matter systems as microscopically distinct as polymers~\cite{struik}, spin glasses~\cite{complexproc}, molecular glasses~\cite{lehenynagel98,agingLunkenheimer}, colloidal gels~\cite{cipgel,gel2}, disordered ferroelectrics~\cite{ferro1,ferro2}, and crumpled paper sheets~\cite{crumpled}. The widespread occurrence of aging phenomena is theoretically understood as a general consequence of frustration leading to a complex free-energy landscape~\cite{bouchaud92,bouchaud1998out}.

Specific experimental protocols, such as temperature cycles, are used to better characterize the nonequilibrium dynamics of glasses~\cite{berthier2002hiking,complexproc}. Temperature cycles within the glass phase were first performed in spin glasses, revealing spectacular dynamical effects~\cite{Lefloch1992,Andersson1993,dupuis,yosh4,yosh3,bert}. Aging is reinitialized after a second downward jump in temperature (rejuvenation), but when the first temperature is restored, the system recalls the state reached before that jump (memory). However, when similar protocols are applied to molecular glasses, such as glycerol, no rejuvenation is observed~\cite{lehenynagel98}, although some memory can be found~\cite{bellon2000,bellon2002advanced,mckenna2003mechanical}.
Aging is a simple consequence of long relaxation timescales, but rejuvenation and memory effects require a specific, hierarchical organization of the free-energy landscape~\cite{hierarchical94,hierarchical95,parisitree,bouchaud2001separation}. This is exactly realized in mean field models for spin-glasses~\cite{MPV87,Pa07b,derridaREM}, and can directly be confirmed in spin-glass simulations~\cite{berthier2002geometrical,berthier2005temperature}.

\begin{figure}[b]
\includegraphics[width= 1. \linewidth]{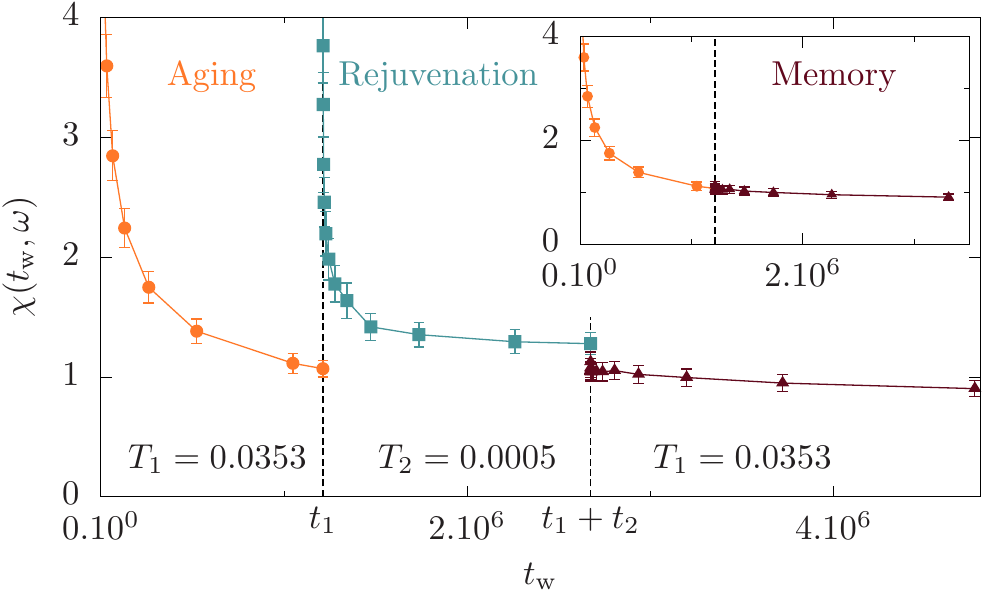}
\caption{{\bf Aging, rejuvenation and memory} in a structural glass submitted to a temperature cycle. We show the time evolution of $\chi(\tw,\omega = 10^{-5})$ in each step of the cycle, delimited by vertical dashed lines. Aging is observed as the liquid is quenched into the glass phase (circles). The glass rejuvenates as it is cooled further (squares), but retains perfect memory when heated back (triangles). In the inset, the intermediate step is removed to better demonstrate the memory.} 
\label{fig:6}
\end{figure}

Recently, the mean-field theory for structural glasses predicted the existence of marginally stable glass phases characterized by a hierarchical free energy landscape, with strong similarities with spin glasses~\cite{CKPUZ16}.
Although the existence of a sharp phase transition between normal and marginally stable glass phases remains debated in finite dimensions~\cite{UB15,CY17,Moore:2017,charbonneau2018morphology}, the theory makes crisp predictions regarding the physical conditions where the glassy landscape becomes hierarchical~\cite{SBZ19}. There are numerical evidences that a complex aging dynamics emerges in the hard sphere model~\cite{BCJPSZ16,SZ18,LB18}, but simulations of model atomic glasses~\cite{Bea:2017,SBZ17} did not find those signatures.

We numerically study the nonequilibrium dynamics of soft repulsive spheres in $d=3$. By carefully choosing the state points where marginal stability is expected~\cite{wca3d} to perform temperature cycles, we successfully observe rejuvenation and memory effects in our model for structural glasses. 
Our central result is presented in Fig.~\ref{fig:6}, where we adopt the same  representation as in experiments, showing the evolution of a dynamic susceptibility $\chi(\tw,\omega)$ (Eq.~\ref{eq:chi}) during the cycle. A high-temperature liquid is rapidly cooled to $T_1$ in the glass phase. Aging dynamics is signaled by a slowly decreasing $\chi$ (circles). The glass is aged for a given time before being cooled to a lower temperature $T_2$. The glass then rejuvenates, since a strong restart of the aging dynamics takes place at $T_2$ (squares). When the glass is reheated to $T_1$ (triangles), it recovers memory of the initial aging (Fig.~\ref{fig:6}, inset), despite the strong rejuvenation in the intermediate step. We attribute these effects to the hierarchical landscape of structural glasses in a marginally stable phase. 
 
\begin{figure}[t]
\includegraphics[scale = .8]{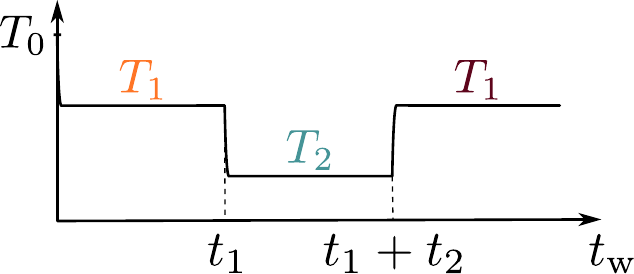}
\caption{{\bf Protocol:} Sketch of the temperature cycle, where $T_1$ and $T_2$ are both in the glass phase while $T_0$ is in the fluid.}
\label{fig:1}
\end{figure}

\textit{Model and methods} --
We study a three-dimensional glass former composed of $N = 3000$ continuously polydisperse particles. Two particles $i$ and $j$ at positions $\bold{r}_i$ and $\bold{r}_j$ interact via the Weeks-Chandler-Andersen (WCA) potential~\cite{WCA71a}
\beq
v(r_{ij}) = 4\epsilon \argc{ ( \sigma_{ij} / r_{ij} )^{12} - (\sigma_{ij} / r_{ij} )^{6} } + 1 \quad,
\eeq
only if they are at a distance $r_{ij} = | \bold{r}_i  - \bold{r}_j|  < 2^{1/6} \sigma_{ij}$, with a non-additive interaction rule $\s_{ij} = \frac{\s_i + \s_j}{2}\argp{1 - 0.2 |\s_i - \s_j|}$. The potential and forces are continuous at the physical cut-off distance.
For each particle, $\s_i$ is drawn from the normalized distribution $P(\sigma_{m} \leq \sigma \leq \sigma_{M}) \sim 1/\sigma^{3}$, where $\sigma_{m} = 0.73$ and $\sigma_{M} = 1.62$. This model is chosen for its excellent glass-forming ability when simulated either with molecular dynamics, or particle-swap dynamics~\cite{NBC17}, and represents a canonical model for dense supercooled liquids~\cite{BT11}.  

The aging dynamics is studied with molecular dynamics (MD). The simulations are performed with a time discretization $dt = 0.003$, within a cubic box of linear size $L$, using periodic boundary conditions. The temperature is controlled by a Berendsen thermostat with damping parameter $\tau_B = 1$~\cite{berendsen}. We reset the total momentum to zero every $10^6$ MD steps. % to avoid accumulation of numerical errors during long MD runs. 
Lengths, times and energies are expressed in units of $\overline{\sigma} = \int  \sigma P(\sigma) d\sigma$, $\sqrt{\eps / m \overline{\sigma}^2}$ and $\epsilon$, respectively. 
The state of the system is determined by temperature $T$, and packing fraction $\varphi = \pi/(3\sqrt{2}L^3) \sum_i \sigma_i^3$. For this non-additive polydisperse mixture, the jamming transition occurs near 
$\varphi_{J} \sim 0.78$. Here, we focus on a fixed packing fraction $\varphi = 0.85$, and discuss later this choice. At this density, the onset of glassy dynamics is near $T_{\rm onset} = 0.2$, at the dynamics slows down by a factor $10^4$ at $T_{\rm c} = 0.07$, below which conventional MD simulations do not reach equilibrium. In addition, we use a hybrid Swap Monte Carlo method~\cite{berthier2018efficient} to prepare equilibrated configurations deep in the glass phase, down to $T = 0.035 \sim T_{\rm c}/2$, to better analyze rejuvenation effects.    

\begin{figure}[t]
\includegraphics[scale = 1.]{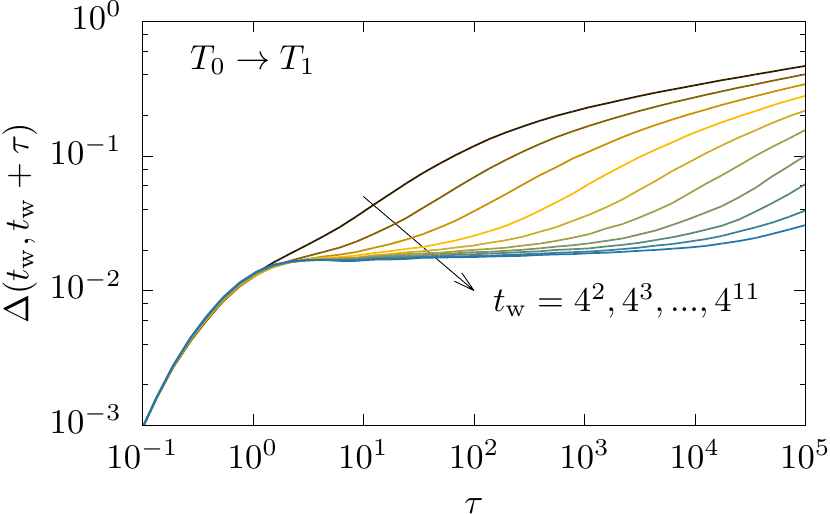}
\caption{ {\bf Aging} of the mean-squared displacement after a quench from the liquid $T_0 = 0.36$ to the glass phase $T_1 = 0.0353$. Each curve corresponds to a given waiting time $\tw$ after the quench, and reveals a  slower motion in older systems.}
\label{fig:2}
\end{figure}

\textit{Protocol and observables} --
We investigate the nonequilibrium dynamics of glasses during a temperature cycle sketched in Fig.~\ref{fig:1}. In the first step, an equilibrium liquid at $T_0 = 0.36$ is quenched rapidly (with a rate of $3.10^{-3}$) to  $T_1 = 0.0353 < T_{\rm c}$. The liquid falls out of equilibrium and slowly ages for a duration $t_1$. In the second step, the aging glass is rapidly cooled to a lower temperature $T_2 < T_1$. It stays there during a time $t_2 = t_1$, after which the system is heated back to $T_1$. 
We measure the mean-squared displacement (MSD): 
\beq
\Delta(\tw,\tw+\tau) =\frac{1}{N} \sum_{i=1}^{N}  \langle |\boldsymbol{r}_i(\tw+\tau) - \boldsymbol{r}_i(\tw)|^2 \rangle \quad,
\label{eq:msd}
\eeq
where $\tw$ is the waiting time after a temperature change. This protocol is repeated using 200 independent equilibrium liquids. The brackets in Eq.~(\ref{eq:msd}) represent an average over these independent runs. 
%We focus on the MSD as it does not explicitly involve a ``probing'' length scale, contrary to the self-intermediate scattering function. 
To make a connection with experiments, we define a dynamic susceptibility~\cite{berthier2002geometrical} 
\beq
\chi(\tw,\omega) = \frac{\Delta(\tw,\tw+\omega^{-1})}{T} \quad,
\label{eq:chi}
\eeq
which plays a role analogous to the ac magnetic or dielectric susceptibility at frequency $\omega$ in experiments. This quantity conveniently compares results at different temperatures, since typical displacements are scaled by $T$, the natural scale for particle motion.

We shall study the role of temperature $T_2$ on the non-equilibrium dynamics of glasses during a temperature cycle as well as the influence of time $t_1$ spent at temperature $T_1$. In particular, we can easily study the limiting case $t_1 \rightarrow \infty$, which corresponds to reaching equilibrium at $T_1$ by generating equilibrium configurations at this temperature using the Swap Monte Carlo method. These very stable glasses would be inaccessible by conventional MD.

\begin{figure}[t]
\includegraphics[scale = 1]{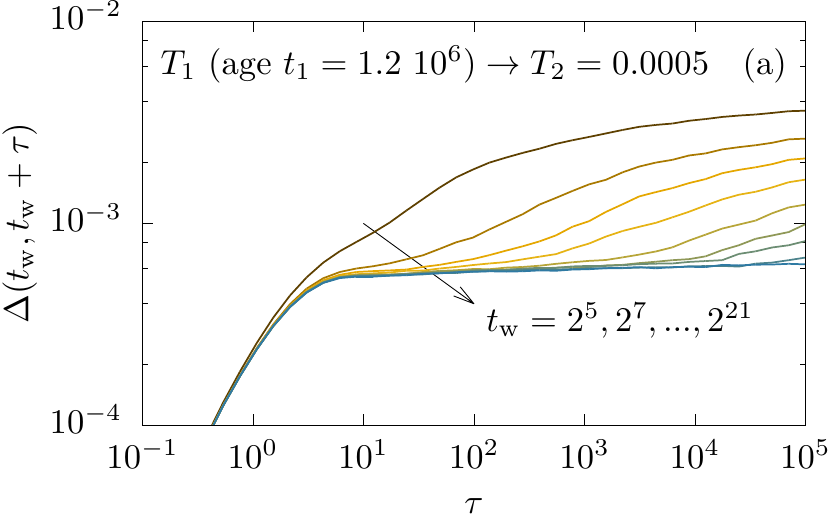}
\includegraphics[scale = 1]{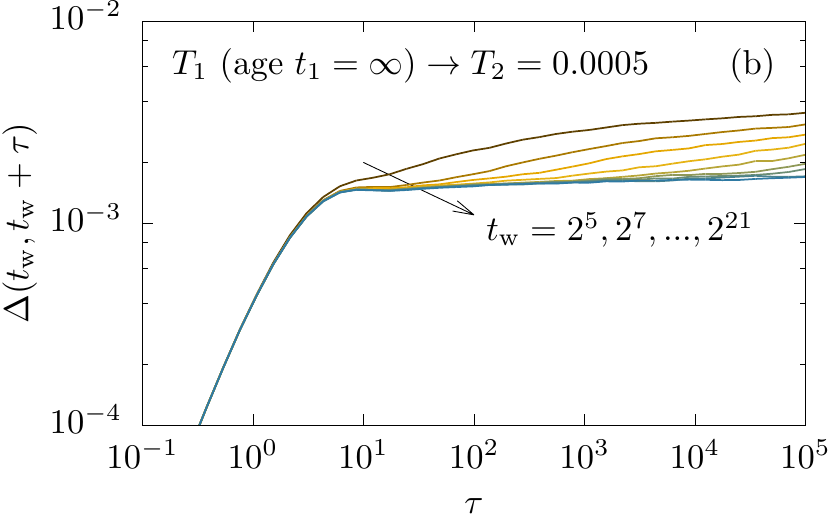}
\caption{{\bf Rejuvenation effect}, or restart of aging dynamics, as glasses aged for (a) $t_1 = 1.2~10^6$ and (b) $t_1 = \infty$ (equilibrium) at $T_1 =0.0353$ are cooled down to $T_2 = 0.0005$. }
\label{fig:3}
\end{figure}

\textit{Aging} -- Let us focus on the first step of the temperature cycle, where liquids thermalized at $T_0 = 0.36$ are rapidly cooled to low temperature, $T_1 = 0.0353 < T_{\rm c}$. The waiting time $\tw$ measures the time spent at $T_1$. The resulting MSD is presented in Fig.~\ref{fig:2}, each curve corresponding to a given waiting time $\tw$. The curves share a similar trend. The MSD increases quadratically at small times $\tau$, before crossing over to a plateau value during a time that depends on $\tw$, and eventually departs from this plateau at larger times. In terms of microscopic dynamics, this corresponds to a short-time ballistic motion, transient trapping within an amorphous cage of neighboring particles, and eventual rearrangement of the cage. Diffusive behavior is not observed within the accessible timescale and particles actually move very little, as the MSD is typically one tenth of particle diameter or less. At the largest $\tw \simeq 4. 10^6$, the MSD plateaus over 4 orders of magnitude in time, meaning that the amorphous structure of the glass remains frozen over very long times. 
We observe a clear waiting-time dependence in the dynamics in Fig.~\ref{fig:2}. The dynamics becomes slower with the age $\tw$ of the system, a typical property of aging systems. This can be seen by plotting the same data as a function of $\tau/\tw$, as the curves collapse when $\tw$ is sufficiently large. The phenomenon described here is widely known and observed in glasses of various materials~\cite{struik,bouchaud92,bouchaudincates}.

Aging can be seen as a consequence of the rugged nature of the landscape of glasses. This corresponds to the thermally-activated crossing of barriers, which leads the system to slowly relax towards lower energy states, where it stays for longer times~\cite{bouchaud92}. The common wisdom in structural glasses is to view these glassy states as energy minima with no (or simple) internal structure~\cite{DS01}, suggesting that no interesting dynamic effect should take place by further cooling the glass. We now present results challenging this view. 

\textit{Rejuvenation} -- We consider the second step of the cycle.
The glasses aged during a time $t_1$ at temperature $T_1$ are suddenly cooled to $T_2 < T_1$. To investigate the influence of $T_2$,  we present data for $T_2 = 0.01$ and $T_2 = 0.0005$. We also consider glasses of two different ages, $t_1 = 1.2~10^6$ and $t_1 = \infty$, the latter being obtained using the hybrid Swap method. As before, we measure the MSD, $\tw$ now being the time spent at $T_2$. 

\begin{figure}[t]
\includegraphics[scale = 1.]{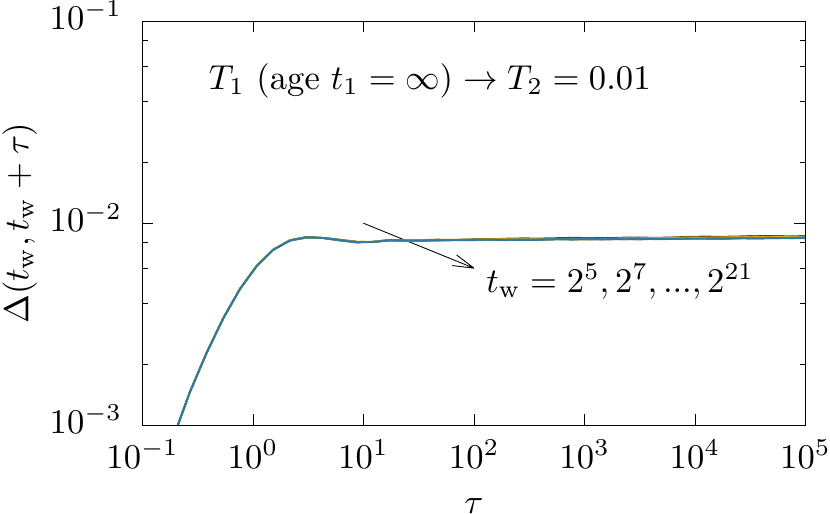}
\caption{{\bf No rejuvenation} if $T_2$ is too large. Here, glasses aged for $t_1 = \infty$ at $T_1 =0.0353$ are cooled down to $T_2 = 0.01$.}
\label{fig:4}
\end{figure}

We start with a large temperature jump to $T_2 = 0.0005$, and report data for 
$t_1= 1.2~10^6$ and $t_1 = \infty$ in Fig.~\ref{fig:3}(a-b).
In both panels, a strongly aging dynamics is observed, similar to the one observed in the first step in Fig.~\ref{fig:2}. The MSD evolves continuously 
over 5 orders of magnitude in time, with strong waiting-time dependence and a variation of one order of magnitude in amplitude. Remarkably, these strong effects survive in Fig.~\ref{fig:3}(b) for $t_1 = \infty$. This implies that the aging dynamics at $T_2$ is not simply the continuation of the one at $T_1$, but that new slow processes emerge at low temperature. This is precisely the rejuvenation effect first reported in spin glasses, since very old glasses (with $t_1=\infty$) age as young glasses at lower temperatures.

Rejuvenation is not observed if $T_2$ is too high. We show in Fig.~\ref{fig:4} the results of cooling from $T_1 = 0.0353$ and $t_1 = \infty$ down to $T_2 = 0.01$. Here, the dynamics does not depend on $\tw$, signaling the absence of aging. The particles are simply trapped in a frozen amorphous structure, which does not evolve further. This is again similar to observations in spin glasses~\cite{bouchaud2001separation,berthier2002geometrical}.
 
\begin{figure}[t]
\includegraphics[scale = 1]{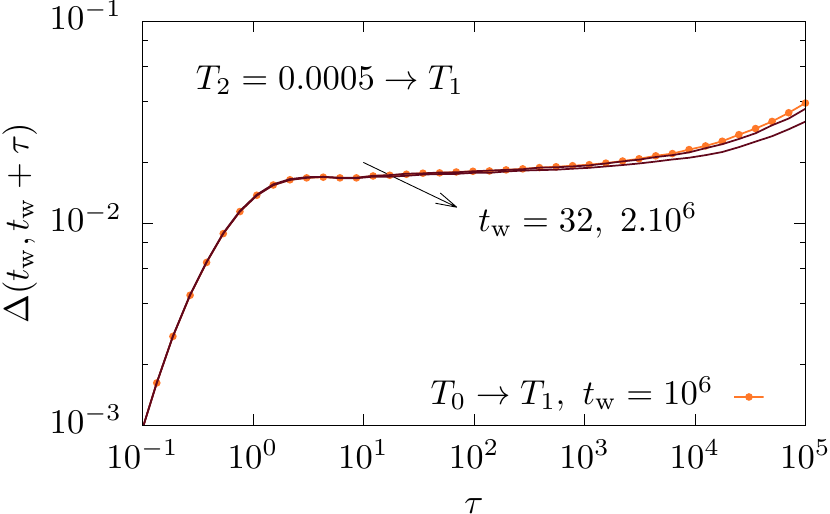}
\caption{{\bf Memory effect} after heating the glass from $T_2 = 0.0353$ (age $t_2= 1.4~10^6$) back to $T_1 = 0.0353$. We report the MSD after heating for $\tw = 32$ and $\tw = 2.10^6$ (lines). The glass has kept memory of its state at temperature $T_1$, as the dynamics smoothly continues that of the first cycle shown for $\tw = 10^6$ (circles).}
\label{fig:5}
\end{figure}

\textit{Memory} -- We complete the thermal cycle by heating the glass from $T_2 = 0.0005$ after $t_2 = 1.4~10^6~\simeq t_1$ back to $T_1 = 0.0353$. The MSD measured after the heating is shown in Fig.~\ref{fig:5}, along with the MSD at the last $\tw$ of the first step of the cycle. After going back to $T_1$, the relaxation dynamics is the direct continuation of the aging which took place in the first step. Despite the strong rejuvenation effect observed at $T_2$ in the intermediate step, the glass has kept a perfect memory of the situation it was in at $T_1$. The aging dynamics then continues as if the second step had not taken place at all. This is the memory effect~\cite{complexproc,berthier2002hiking}. 

We gather all these results in Fig.~\ref{fig:6} by reporting the time evolution of $\chi(\tw,\omega=10^{-5})$, defined in Eq.~\ref{eq:chi}, during the complete temperature cycle. Aging in the first part of the cycle corresponds to a slow decay of $\chi(\tw,\omega)$, while rejuvenation corresponds to a strong restart of a similar aging. Memory is very clear in the third step which appears as the direct continuation of the first one, as emphasized in the inset where the second step is removed. The aging dynamics in the third step proceeds as a simple continuation of the first.
This figure mirrors similar results obtained in spin glass materials~\cite{refregier1987ageing,complexproc,berthier2002geometrical,berthier2005temperature}.
The simultaneous observation of both rejuvenation and memory effects is highly non-trivial, and supports the idea that the landscape inside glassy minima can be rugged and hierarchical in systems of soft repulsive particles that describe structural glasses.

\textit{Separation of lengthscales} -- We have studied the probability distribution function (pdf) of single particle displacements in the three steps of the temperature protocol. At each temperature, we measure the pdf of  $\Delta r = |\boldsymbol{r}(\tw+\tau)  - \boldsymbol{r}(\tw) |^2$ towards the end of the step, for $\tw = 2^{20}$ and $\tau = 10^5$. These distributions give additional information on the typical scale of particle displacements at each temperature (the average value is plotted in Figs.~\ref{fig:2}-\ref{fig:5}), and, more importantly on the heterogeneity of the particle displacements.

\begin{figure}[t]
\includegraphics[width= \linewidth]{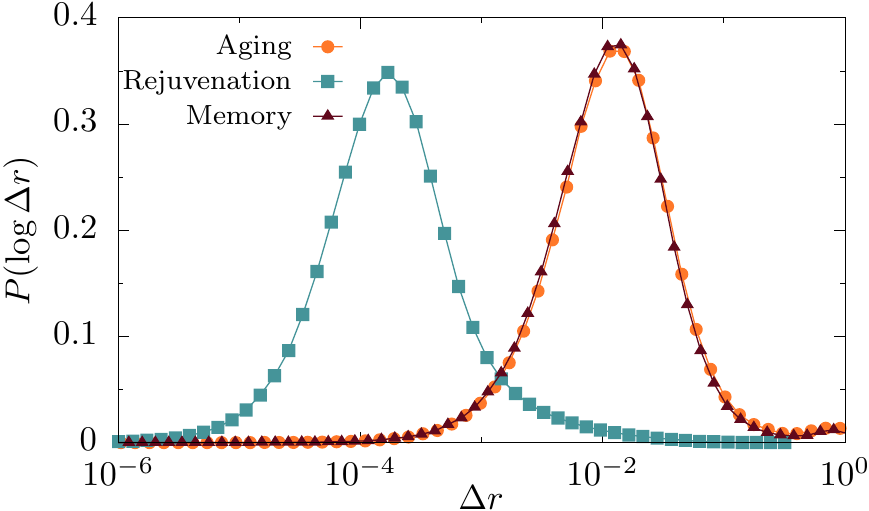}
\caption{{\bf Hierarchy of lengthscales} in the probability distribution function of particle displacements $\Delta r$ during aging (circle), rejuvenation (square), and memory (triangle) for $\tw=2^{20}$ and $\tau=10^5$. All particles contribute collectively to each step, but at a different scales.}
\label{fig:7}
\end{figure}

Results for the three steps are reported in Fig.~\ref{fig:7}. 
We observe that the pdf of displacements during aging is broad but relatively featureless, indicating that all particles are involved in the aging dynamics. 
A similar shape is obtained during the rejuvenation, but at a much smaller scale. This indicates that the aging dynamics in the second step is again due to very collective particle motion involving the entire system, but it involves displacements at a much smaller lengthscale. This explains why memory of the first step is retained, as the structure obtained at the end of the first step is essentially unperturbed during the second step. Dynamics is hierarchical both in timescales and in lengthscales~\cite{bouchaud2001separation,bouchaudincates}.   

%and rejuvenation have a comparable width. This means that both aging and rejuvenation processes involve all particles in the sample. The average displacement is naturally much smaller during rejuvenation, which takes place at $T_2 \ll T_1$. As the temperature is shifted back to $T_1$, not only the average displacement at $T_1$ is remembered (see Fig.~\ref{fig:6}), but the complete pdf is recovered. The pdf of displacements in the final memory step is the same as during aging, slightly shifted to lower values. We see clearly in Fig.~\ref{fig:7} that the dynamical processes taking place at $T_1$ and $T_2$ occur on well-separated lengthscales. The rejuvenation effect at $T_2$ is distinct from the aging process taking place at $T_1$, as it involves relaxations on much smaller lengthscales. Because of this separation of lengthscales, the strong rejuvenation effect at $T_2$ does not affect the relaxation processes occurring at $T_1$ in which a memory of the age of the glass is encoded.

\textit{Conclusion} -- We have shown that submitting a three-dimensional model for structural glasses to a temperature cycle reveals rich non-equilibrium dynamical effects, such as rejuvenation and memory effects that were first observed in spin glasses, but not in molecular glasses.  
Performing an extensive study of the phase diagram of the WCA model analyzed in this work, we have found that for $\varphi = 0.85$ and $T_1=0.0353$, only quenches below $T_2 \approx 0.001$ will lead to rejuvenation effects in the dynamics. We have also analyzed the density dependence of these effects and found that no such rejuvenation effect can be found for packing fractions beyond $\varphi \approx 0.9$. These findings, which will be reported in detail elsewhere~\cite{wca3d}, suggest that soft repulsive spheres at packing fractions relevant to describe soft colloids and granular materials should give rise to rejuvenation and memory effects, whereas in the regime describing dense supercooled liquids they will not. These conclusions are broadly consistent with mean-field analysis~\cite{SBZ19}, they can explain the absence of rejuvenation reported for glycerol~\cite{lehenynagel98}, and should guide future experimental studies of the dynamics of glassy materials. 

\vspace{.3cm}
We thank F. Zamponi, Q. Liao, H. Yoshino and S. R. Nagel for useful exchanges. This work was supported by a grant from the Simons Foundation (\#454933, L. Berthier).

\bibliographystyle{mioaps}
\bibliography{memory}

\end{document}